\documentclass[conference]{IEEEtran}
\usepackage{listings}
\usepackage{setspace}

\renewcommand{\baselinestretch}{.948} 

\usepackage{fancyhdr}
\usepackage{float}
\usepackage{xcolor}

\definecolor{codegreen}{rgb}{0,0.6,0}
\definecolor{codegray}{rgb}{0.5,0.5,0.5}
\definecolor{codepurple}{rgb}{0.58,0,0.82}
\definecolor{backcolour}{rgb}{0.95,0.95,0.92}

\lstdefinestyle{mystyle}{
    backgroundcolor=\color{backcolour},   
    commentstyle=\color{codegreen},
    keywordstyle=\color{magenta},
    numberstyle=\tiny\color{codegray},
    stringstyle=\color{codepurple},
    basicstyle=\ttfamily\footnotesize,
    breakatwhitespace=false,         
    breaklines=true,                 
    captionpos=b,                    
    keepspaces=true,                 
    numbers=left,                    
    numbersep=5pt,                  
    showspaces=false,                
    showstringspaces=false,
    showtabs=false,                  
    tabsize=2
}

\ifCLASSINFOpdf
  \usepackage[pdftex]{graphicx}
\else
\fi

\begin{document}

\title{SDFLMQ: A Semi-Decentralized Federated Learning Framework over MQTT}

\author{
\IEEEauthorblockN{Amir Ali-Pour}
\IEEEauthorblockA{\textit{Department of Software and IT Engineering} \\
\textit{ÉTS Montréal / Université du Québec}\\
Montreal, Quebec, Canada \\
amir.ali-pour@etsmtl.ca}
\and
\IEEEauthorblockN{Julien Gascon-Samson}
\IEEEauthorblockA{\textit{Department of Software and IT Engineering} \\
\textit{ÉTS Montréal / Université du Québec}\\
Montreal, Quebec, Canada \\
julien.gascon-samson@etsmtl.ca}
\thanks{}}


\maketitle

\begin{abstract}
Federated Learning is widely discussed as a distributed machine learning concept with stress on preserving data privacy. Various structures of Federated Learning were proposed. Centralized Federated learning for instance has been the primary structure that suits cloud computing. Decentralized Federated learning also has been proposed for ecosystems where communication is dominantly peer-to-peer. Semi-Decentralized Federated Learning (SDFL) has emerged recently as a new concept where the interconnected nodes are clustered, and each cluster is managed independently. The potential of SDFL lies in its clustering feature, which distributes the load of the global model update down onto multiple nodes. Since the concept is fairly new, much can be done to render this FL model a reliable, efficient, and real-time service at the edge. In this paper, we propose SDFLMQ, a semi-decentralized Federated learning framework at the Edge that uses MQTT as the communication protocol. We demonstrate how the publish/subscribe communication model is used to facilitate the clustering and load balancing in SDFL. We also demonstrate how SDFLMQ can use some of the core MQTT features to expand its capacity with no significant costs. Based on some primary evaluations, we demonstrate how SDFLMQ can efficiently distribute the load of aggregation, and potentially save unnecessary memory allocation, all with no requirement for a powerful central unit for aggregation and global model update. We also disclose some of the key future expansions of SDFLMQ with a focus on the operation of large deep neural network models at the edge. 
\end{abstract}

\begin{IEEEkeywords}
Federated Learning, Distributed Systems, Edge Computing, MQTT, Publish/Subscribe Communication, IoT.
\end{IEEEkeywords}

\IEEEpeerreviewmaketitle

\begin{figure*}[t]
    \centering
    \includegraphics[width=1\linewidth]{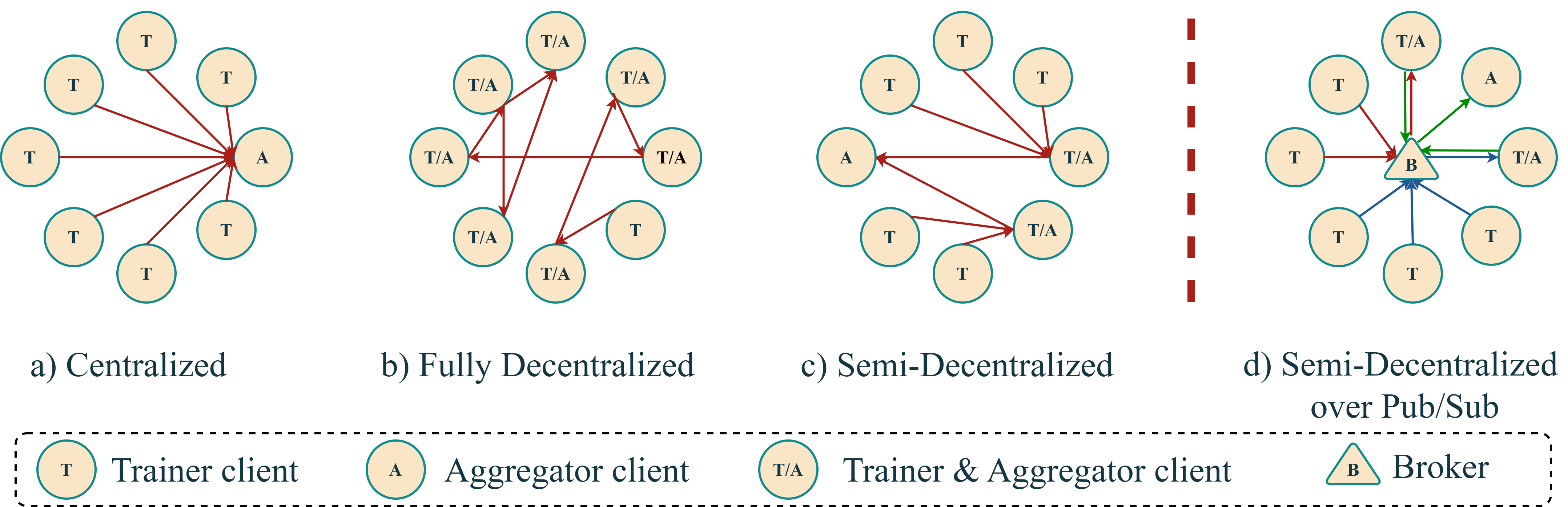}
    \caption{Federated Learning Topologies}
    \label{fig:fl_topologies}
\end{figure*}

\section{Introduction}

Federated Learning (FL) has emerged as a transformative approach for distributed machine learning in Internet of Things (IoT) ecosystems  \cite{nguyen2021federated, zhang2022federated}. The proliferation of IoT devices has generated vast amounts of decentralized data at the network edge, presenting significant challenges for traditional centralized learning paradigms. These conventional approaches necessitate the transfer of data to centralized servers, which incurs considerable bandwidth costs, exacerbates latency and poses serious privacy risks. In contrast, FL enables collaborative model training directly on edge devices, allowing them to update a shared global model without exchanging raw data \cite{lim2020federated}. This characteristic is especially pertinent for IoT systems, where efficient bandwidth usage, enhanced privacy, and real-time response capabilities are critical to system performance. Thus, FL addresses key limitations in centralized learning by reducing data transmission overheads, preserving user privacy, and optimizing computational resources at the edge \cite{ji2023joint, guo2021efficient}.

Federated learning can be categorized into three principal topologies: centralized, fully decentralized, and semi-decentralized, each offering distinct architectural frameworks \cite{bonawitz2019towards, beltran2023decentralized}. Centralized FL relies on a single server to coordinate the model aggregation process. Participating devices perform local training and send their updates to the central server, which aggregates them to form a global model, subsequently disseminated to the devices. Conversely, fully decentralized FL eliminates the need for a central coordinating entity, with devices directly communicating and exchanging model updates in a peer-to-peer manner. This topology ensures no single point of failure but introduces complexities in synchronization and network efficiency. Semi-decentralized FL represents an intermediary approach, wherein clusters of devices are organized, each managed by a local coordinator. These local coordinators aggregate the updates within their clusters and coordinate with a central server or upper-tier coordinator for global aggregation, thus blending decentralized communication within clusters and centralized oversight across the network.

Each federated learning topology exhibits distinct advantages and limitations, particularly when deployed in heterogeneous IoT environments. Centralized FL, while relatively simple in its coordination, suffers from scalability bottlenecks and single points of failure. This approach is often unsuitable for large-scale IoT systems where robustness and network scalability are paramount. On the other hand, fully decentralized FL mitigates these concerns by removing reliance on a central server. However, this topology incurs significant overhead in inter-device communication and poses challenges related to the consistency of the global model due to varying update frequencies and network conditions. In contrast, semi-decentralized FL offers a more scalable and resilient framework by distributing coordination responsibilities among local aggregators while preserving overarching global model integrity \cite{xu2024edge, luo2020hfel}. This topology is particularly beneficial for heterogeneous IoT systems, where devices exhibit diverse computational capabilities and data distributions. By enabling local aggregation within clusters, semi-decentralized FL reduces communication overhead and improves fault tolerance, making it well-suited to large-scale, IoT networks. For both centralized and fully decentralized FL topologies, established platforms such as PySyft \cite{ziller2021pysyft}, FedML \cite{he2020fedml}, and Flower \cite{beutel2020flower} provide comprehensive frameworks to facilitate distributed model training. These platforms support a wide range of applications by offering APIs and tools for coordinating both centralized and peer-to-peer federated learning processes. Despite this maturity in platform development, there is a notable absence of dedicated platforms for semi-decentralized FL. Existing solutions are not fully optimized to handle the hierarchical and clustered structure inherent to semi-decentralized architectures, where communication and coordination must be efficiently managed within clusters while maintaining global model synchronization. This gap in the current landscape underscores the need for specialized platforms that can address the unique demands of semi-decentralized FL, particularly for heterogeneous IoT systems with variable node configurations.
In this paper, we propose a framework that utilizes the topic-based communication model in Publish/Subscribe communication protocols to perform dynamic clustering and balance the load of model aggregation over several contributing clients. With our proposed SDFLMQ framework, we show how a group of inter-connected nodes through MQTT communication protocol can perform both local and global model updating in synchronization, elevating the need for a central server with excessive cost to perform the tasks. 

The flow of the paper would be the following: In section II we elaborate on the motivation behind designing and developing SDFLMQ. In section III we discuss the architecture and details on the logic of SDFLMQ framework and it's constituent components. In section IV we briefly discuss the implementation of SDFLMQ in Python, and in section V we demonstrate a potential use-case of the framework. In section VI also, we demonstrate some run-time evaluation of SDFLMQ. In section VII we elaborate on some of the key future steps in expanding the framework, and in section VIII we conclude the paper.

\begin{figure*}[t]
    \centering
    \includegraphics[width=1\linewidth]{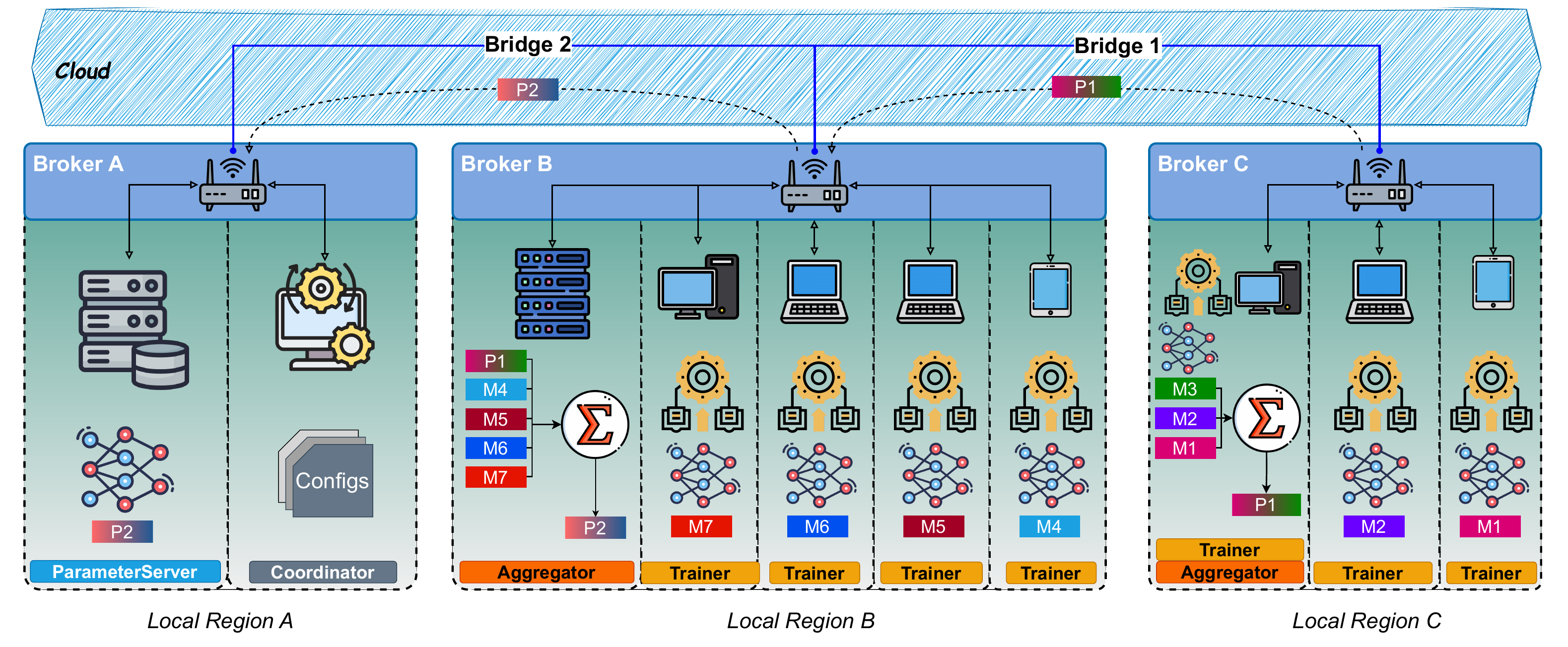}
    \caption{overview of semi-decentralized federated learning (SDFL) over MQTT communication with broker bridging enabled.}
    \label{fig:sdfl_overview}
\end{figure*}

\section{Motivation}

The conventional way of performing FL is to consider a central point of aggregation. This model is trivial and commonplace in many industrial applications. However, when a massive set of devices aims to contribute to the FL process, the single aggregation point becomes a bottleneck. Therefore, it can cause connection failure due to network congestion, bandwidth, and/or memory overflow. That is why semi-decentralized federated learning is proposed, where clustering is introduced and mid-level aggregation is performed in each cluster before it reaches the final point of aggregation. In this way, the aggregation load is distributed, thus the probability of network congestion and local memory overflow are reduced. Fig \ref{fig:fl_topologies} shows different topologies of federated learning.

The common communication model for existing Semi-decentralized FL is the client/server model with fixed aggregator placement. For instance, in the case of mobile devices, base stations are the mid-level aggregators before the final level of aggregation \cite{sun2023semi}. In such an establishment, the base station is a fixed point, and mobile devices communicate to the suitable base station to send their local model update and receive the global model updates.

For such establishments where a powerful system can be considered that has a large memory capacity and a stable network connection, of course, a client/server communication model can be considered suitable. However, there could be other scenarios, wherein the point of mid-level aggregation is not a powerful system like that in a base station. We can imagine a scenario for IoT devices in an enclosed interconnected environment, where all the connected systems have computation or power constraints. If we decide to place SDFL over such establishment, we need to i) be able to assign the role of aggregation to existing devices with resource constraints, and ii) have a format of dynamicity that allows us to change the role of mid-aggregation from round to round to avoid device exhaustion and overloading. This would allow us to adapt to the individual system constraints. However, if the client/server communication model is used, then to implement this dynamic role management, a complex method should be implemented to inform systems in each cluster of who should be the aggregator in the new round. Alternatively, a fully decentralized FL over P2P model can be implemented for such a situation \cite{beltran2024fedstellar}, which can ensure that at least for the role of aggregation, there will be no memory or bandwidth overflow. However, that could come at a cost of extra time for training/aggregation due to the sequential communication. 

In contrast to the client/server communication and fully decentralized FL with a P2P model, the publish/subscribe communication model can be imagined, where systems' roles in the FL process are associated with topics. In this way, if a system is added or removed from the cluster, or subject to changing its role (e.g., from trainer to aggregator or vice versa), that system alone needs to undergo an update, while other systems preserve their connection/topics they have subscribed to or want to publish to. We anticipated that such a mechanism facilitates/is suitable for dynamic role management in semi-decentralized federated learning where aggregation roles need to change frequently due to system constraints. This may be slower compared to the semi-decentralized FL over client/server with fixed aggregation points since all the messages have to go through a broker system, but faster compared to fully decentralized over P2P since there is some level of parallelism in the aggregation process. Also, the model simplifies dynamic role management compared to the Semi-decentralized FL over the client/server with dynamic aggregation placement. Fig \ref{fig:sdfl_overview} shows an overview of how SDFL can operate over MQTT-based communication.

We could also add that such establishment of SDFL over pub/sub can be easy to integrate in a cloud-to-edge continuum. To integrate such a service, we only need a broker at the edge to disseminate the model updates, while the FL-specific roles are managed on the devices that need the ML services. Therefore, at the edge, role association would be as general as just a message disseminator which does not need any adaptation to the FL process. For instance, if an MQTT broker is running as a service on an edge server, we can connect to that and establish the FL roles among the devices connected to the broker. This would in turn help set up the framework faster and with reduced cost of installment.





\begin{figure*}[t]
    \centering
    \includegraphics[width=1\linewidth]{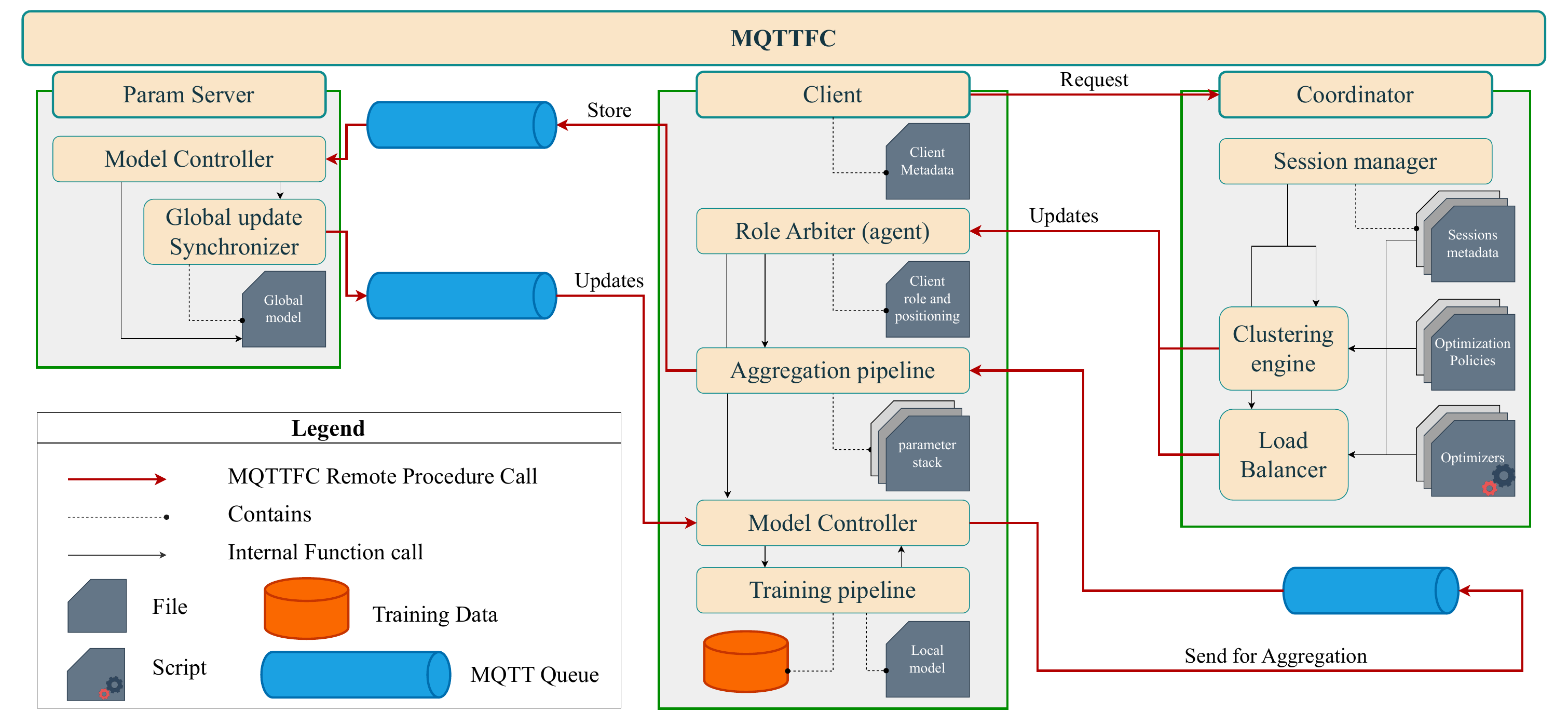}
    \caption{SDFLMQ framework architecture. Both the aggregation and training services are included in the client model. However, based on the role arbiter, either one, or both of them can run and contribute to the FL round. The clustering engine and Task manager services on the coordinator model communicate with the role arbiter on the client system to choose their aggregation and training roles. The aggregation pipeline running on the root node will store the updated model on the parameter server. The global update synchronizer on the parameter server updates the local model in each client after an FL round.}
    \label{fig:sdfl_arch}
\end{figure*}

\section{SDFLMQ Framework}
\subsection{Framework Design Goals}
Unlike the existing FL platforms, SDFLMQ aims to pay more attention to the system constraints and adaptability toward a variety of applications and methodologies. It also is aimed for simplicity in design, thus easy to understand, and tailored for more application-specific purposes. SDFLMQ follows 5 key design goals which are drafted in the following.

\subsubsection{Adaptable to System Constraints}
Role management in SDFLMQ is set to be dynamic. In that, SDFLMQ is designed in a way to read the system constraints and rearrange the playing roles in the FL ecosystem so that certain goals are met aside from achieving high accuracy in the model under training. For instance, a goal would be set to optimize the role arrangement at each round so that the global FL update converges faster to an optimal point, or converge with less resource consumption. The initial roles that the base design of SDFLMQ will be able to manage would be training and aggregation.

\subsubsection{Scalable to Large Number of Clients}
Since the point of aggregation which has a key role in global model updating is distributed, SDFLMQ would be able to support connecting a large number of clients to contribute to the FL process. Client clustering and Broker bridging are some of the key networking features that SDFLMQ integrates to provide a seamless contribution for a large group of clients in training a mutual model.  

\subsubsection{Modular for Further Expandability}
The role association in SDFLMQ would follow some policies that can potentially read the system parameters to optimize the placement or operate as a black-box optimizer without requiring to monitor clients' operational status. Per application or methodology demands, more sophisticated policies need to be developed or integrated which could need more than just the system parameters to associate roles to the contributing clients. The role management policies in this regard are designed to be modular and thus new policies can be added and later used for more specific use-cases. Moreover, other roles may be integrated in the future, to provide support for a more sophisticated FL ecosystem. These roles would be beyond the acts of aggregation and training. For instance, feature extraction, quantization, knowledge distillation, and parameter encoding may be needed which may not be able to run on a singular machine. But, if integrated into SDFLMQ, could be managed in a distributed manner.

\subsubsection{Flexible to Variety of ML/FL Methodologies}
We have a variety of ML models and training strategies that we need to be flexible for in terms of general applications. We also have a variety of FL strategies, such as layer-wise federated learning, federated transfer learning, etc.

\subsubsection{Lightweight Core Functionality}
While the modularity vows for further expansion of the framework, we want to keep the core of the framework simple, and thus easy to integrate. This means that we should be able to easily invoke global model updating by incorporating a few lines of code. We elaborate more on this in section V where we discuss a potential use-case with a code snippet in Python.

\subsection{Architecture} 
\subsubsection{MQTT Fleet Control}

SDFLMQ is based on a tailor-made remote function call (RFC) infrastructure called MQTT Fleet Control (MQTTFC). This lightweight RFC infrastructure simply binds clients' remotely executable functions to MQTT topics. Thus, any remote client can publish to the function topic and pass the arguments within the message payload, and the function will be called in the client system which has the corresponding function and has subscribed to the topic of that function.

\subsubsection{SDFLMQ Components}
The main components of SDFLMQ framework are the coordinator logic, the client logic, and a parameter server logic. These are the building blocks of the framework that provide the key roles including role management and clustering by the coordinator, and training and aggregation by the client. Fig. \ref{fig:sdfl_arch} shows the system architecture of SDFLMQ based on MQTTFC, MQTT, and Python. 

The components run independently and use the MQTTFC RFC mechanism for communication. The Coordinator only receives the metadata needed to perform role arrangement and rearrangement and sends only routing and task placement metadata to the clients. The Client logic represents 1 main class of functions which is the Aggregation. This class includes various techniques to process global model updates based on individual model parameters. In addition, there are peripheral classes including Role Arbiter and Model Controller. The Role Arbiter mainly governs the aggregation task. It indicates whether a client should accept any incoming model parameters if it is an aggregator, and to which node it should send the aggregated model parameter. The Model Controller class also governs the Model repository according to the session they are bound to. Each update that is applied to the model, either locally or globally, updates the Model Controller as well. A Training pipeline would have access to the Model Controller to send and receive model parameters when needed. While the Model Controller in the Client Logic is provided to keep track of the Models, it would only keep track of the models designated for sessions that the corresponding client machine is a part of. To keep track of all the Models associated with all the sessions that are handled by the Coordinator, a Parameter Server Logic is defined. This Logic can run either on the same system that runs the Coordinator or on a separate system. The Parameter Server would listen to a public topic that is designated for sending and receiving Global models. Thus, it serves as a repository for global models.



\begin{figure}[t]
    \centering
    \includegraphics[width=1\linewidth]{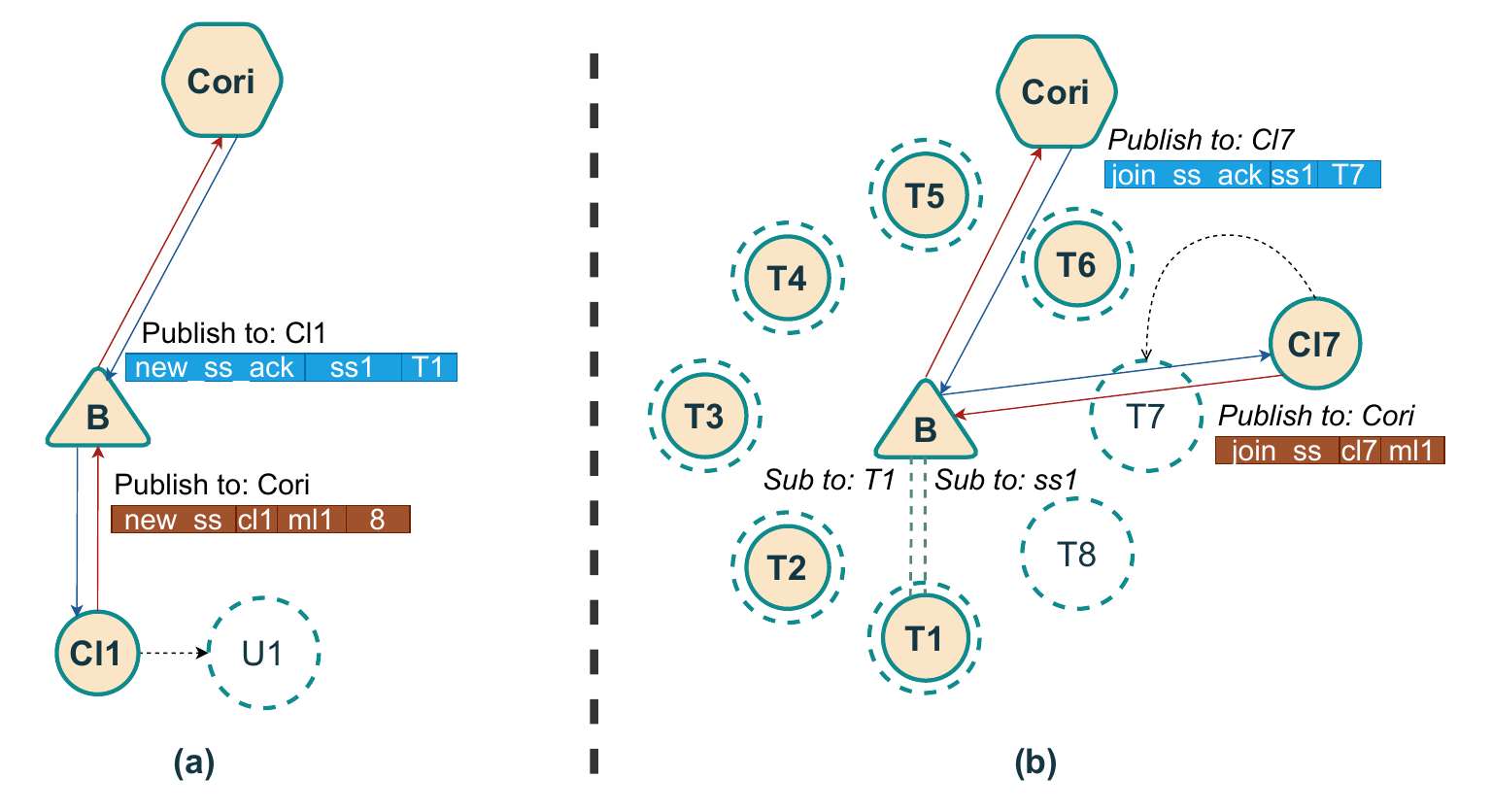}
    \caption{Illustration showing a) interaction between client and coordinator for a new session establishment for up to 8 contributors, and b) interaction between a new client and coordinator for joining a session.}
    \label{fig:coord_1}
\end{figure}

\subsection{Client Logic}
Here we elaborate on how a client can contribute to the global model updating process through SDFLMQ. Generally, a client can hold one of the following roles: 1) Trainer, 2)Aggregator, 3)Trainer/Aggregator. In the following the logic of each role is explained.

\subsubsection{Trainer}
Trainer clients are those who contribute solely by training a designated model locally. The trainer client first updates the model through multiple local rounds (epochs) of training. Then, it enters a status of sending weights to the aggregator and waits for the new global model. 

\subsubsection{Trainer-Aggregator}
Trainer-Aggregator clients can contribute both as a trainer and an aggregator. To do so, it first notifies the coordinator of its preference of being an aggregator. Then, during each federated learning round, it will perform its local training first, then will enter a state to receive model parameters from its preceding nodes. Note that the number of model parameters an aggregator unit should receive is defined by the coordinator after the coordinator has defined the roles and clusters for a session. We will explain this part more in the next sub-section where we explain the client logic.

\subsubsection{Aggregator}
A client can also contribute solely as an aggregator, without needing to collaborate as a trainer. To do so, it sends its preference to take the role as an aggregator, and the coordinator will decide whether the client is suitable for the role or not. For these units, the only task at each federated learning round is the aggregation.  

\begin{figure}[t]
    \centering
    \includegraphics[width=1\linewidth]{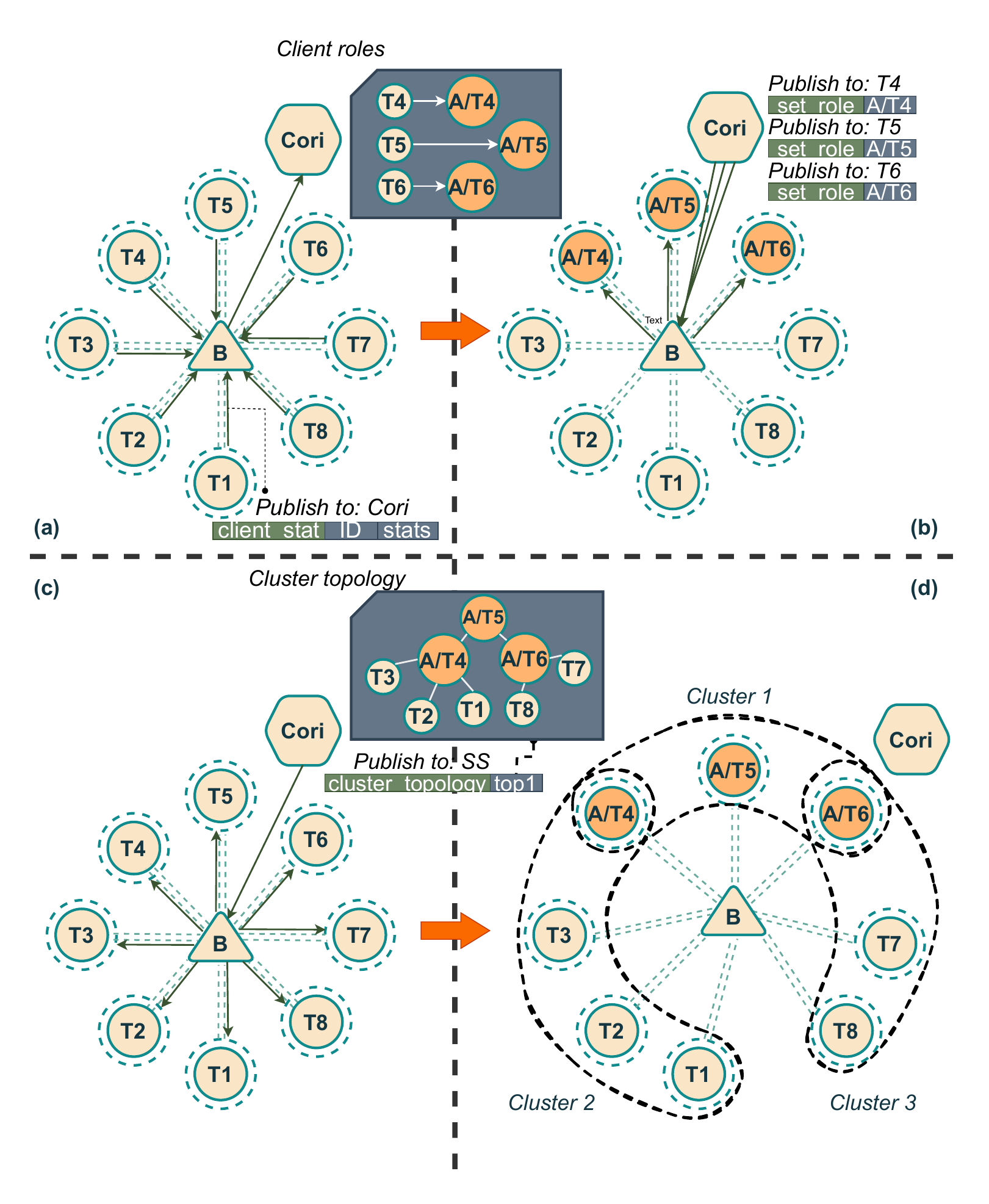}
    \caption{Illustration showing a) interaction between clients and the coordinator, where the coordinator is publishing clustering topology on session topic to all clients b) A clustered network of clients, where A/T5 is the head of cluster 1, A/T4 is the head of cluster 2, and A/T6 is the head of cluster 3.}
    \label{fig:coord_2}
\end{figure}

\subsection{Coordinator Logic}
The coordinator is mainly responsible for governing the aggregation. It is initially based on the MQTTFC executable class and comprises functions for role association, clustering, and session management. 

\subsection{Topic-based Role Management}
The main idea of topic-based role management is to associate the functions of aggregation and training with topics. While the training can run independently of the MQTTFC RFC mechanism, the synchronized global model updates need the aggregation function in the client systems to be called remotely via the coordinator. The coordinator has a role initialization process in which it informs all the clients of their roles and clusters at the beginning of an FL session. It also has a role rearrangement process during which it informs only the clients whose roles and clusters need to change. In the following, the initialization process for session generation, role association, and clustering, as well as role rearrangement are elaborated.


\subsubsection{FL Session Initialization}
A session is created when a new model demands global updating. This demand can be issued in terms of creating a new federated learning session from one of the clients, who will send the request for session initialization for the given model. The model name and client ID are taken from the request message payload alongside other information including the session period, and maximum number of contributors, as shown in Fig. \ref{fig:coord_1}(a), and the session is created. A session will track the model global updates and the contributing clients at each FL round. If two clients send initiation requests, the coordinator will serve the first request, and dump the other one. Once the session is created, the coordinator will wait for contributors to send their request to join the session. To join the session, contributing clients will only need to send their client ID and model ID. Once enough contributors have joined the session, the coordinator will undergo the process of clustering and role arrangement. The session will terminate once it has reached its termination time, or the federated learning round counter has reached its maximum number which is designated at the session initialization phase. Fig. \ref{fig:coord_1}(b) shows the interaction between the coordinator and one of the clients requesting to join a session.

\subsubsection{Clustering}
For clustering, the client chooses first the aggregators. The aggregators are identified as cluster heads also. Once the aggregators are selected, then the trainers are associated with the aggregators. Clients could be both an aggregator and a trainer, given that the clustering is hierarchical. Note that a hierarchy of clusters can exist, wherein the aggregator/trainer clients are at one level the cluster heads (the aggregator of that cluster), and at the higher levels are trainers that contribute to the higher level cluster head. Such hierarchical aggregation distributes the aggregation load temporally, given multiple levels of aggregation, and spatially, given multiple aggregators at each level. Fig. \ref{fig:coord_2} shows the interaction between the coordinator and the contributing clients in forming a hierarchical cluster in SDFLMQ.

\begin{figure}[t]
    \centering
    \includegraphics[width=1\linewidth]{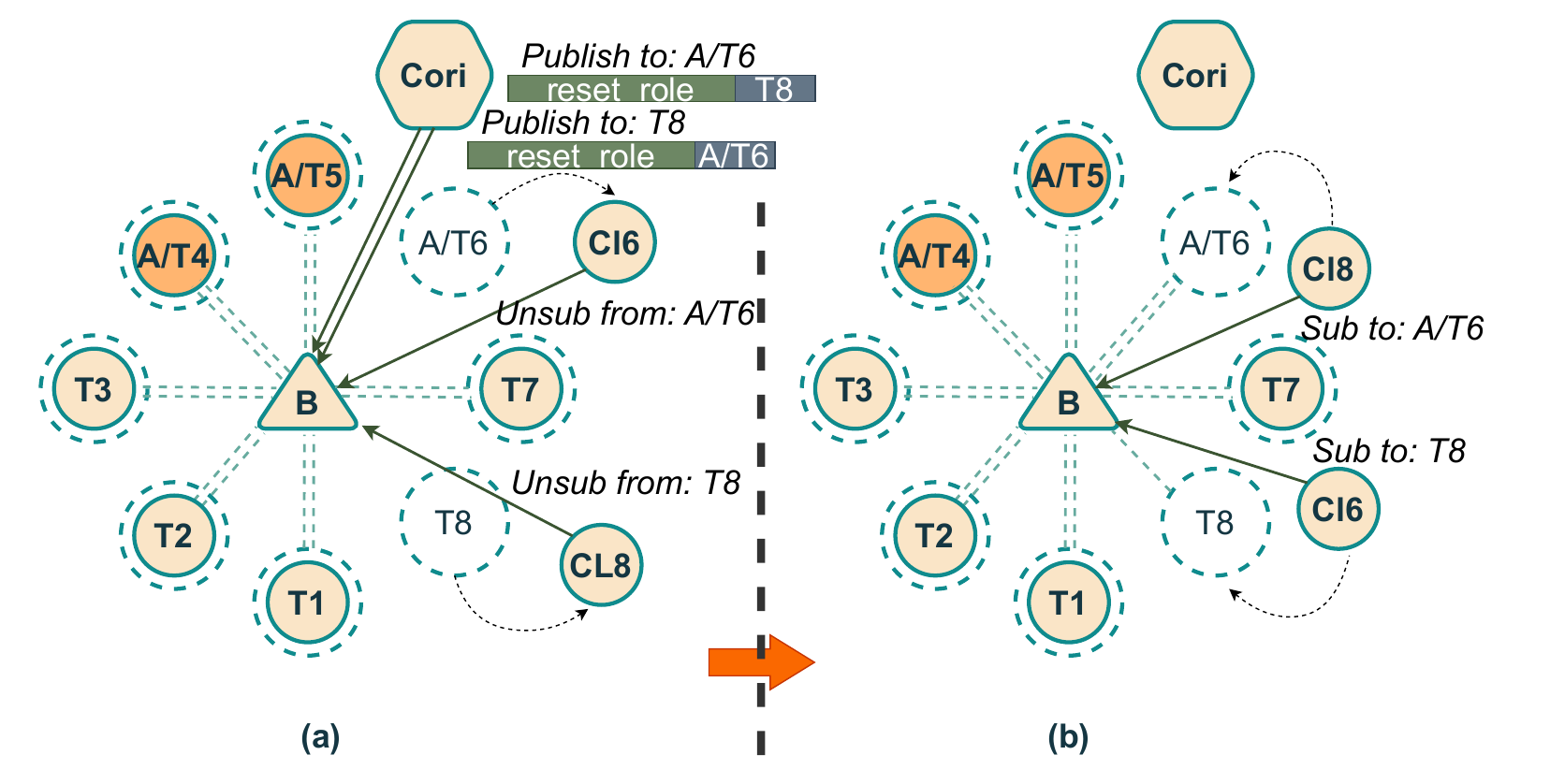}
    \caption{Illustration showing the coordinator interaction with 2 clients switching roles through the role re-arrangement process. In (a) the clients unsubscribe from their current role, and in (b), according to the new role received from the coordinator, they will subscribe to the new roles. Noting that Cl6 will need to update its cluster head from A/T5 to A/T6 in its new role as T8.}
    \label{fig:coord_3}
\end{figure}

\subsubsection{Role Arrangement}
The initial role arrangement is performed during the clustering process as discussed above. The process includes selecting the aggregator, trainer/aggregator, and trainer clients. Note that these are the primary roles in the SDFLMQ ecosystem as mentioned before. Once the coordinator has established a hierarchical cluster model for the session, the process of role arrangement begins. Then, given that each role has a fixed topic associated with it, the coordinator informs the clients through a private channel, of the role topic they need to subscribe and publish to. 

\subsubsection{Session Status Updates}
This logic simply tracks the session status throughout FL rounds. The updates happen after each client has performed its local process and the contribution according to its associated role. Once the client's job is finished, it sends its readiness for the next round, including system stats and model performance if needed, to the coordinator. The coordinator will use these parameters then for role optimization and rearrangement.

\subsubsection{Role Rearrangement}
The process of role rearrangement happens in response to the role optimization output, which changes the roles of clients according to their systematic or data parameters changing throughout the session. Unlike the role arrangement process, however, this process informs only the clients whose roles have changed for the new FL round. Fig. \ref{fig:coord_3} shows the process of role rearrangement, wherein the coordinator informs the selected clients to change their roles according to their stats.

\subsubsection{Role Optimization}
Per se, the clustering and hierarchy of role arrangement in SDFLMQ parallelize the aggregation tasks to avoid overloading model parameters in one system. However, contributing systems' characteristics could change over time, which can affect the overall performance of the FL operation. For instance, the memory capacity of the contributing aggregator client machines can change, in some cases, be lowered, and thus can leave less memory capacity for the receiving model parameters. In such a case, if the machine does not delegate its role to another client with more memory, then the memory overflow can further delay the learning process due to extra load/store from the storage unit, or in some cases lead to local system failure. 

To make the role management logic adaptive to the possible changes in the interconnected system, a load balancer component is implemented for the coordinator, which in turn processes and rearranges the roles of contributing clients periodically. Different optimizers can be used for this purpose. The optimization process repeats per round and chooses the appropriate aggregators for each level of the hierarchy. Noting that the optimizer is a modular logic. This means that depending on the needs of the application, different optimizers can be employed. For instance, one optimizer would process the merits of the clients based only on their systematic characteristics. Another optimizer could process the system parameters as well as the data distribution and model performance. The optimizer plays a key role in the coordinator's logic.

\subsection{Broker Bridging}
One of the key challenges of basing SDFL over a publish/subscribe protocol such as MQTT is that in cases where the number of contributors is high and/or the model parameter size is large, the broker at the center can become a bottleneck. However, MQTT in itself includes a feature called broker bridging, which allows multiple brokers to share their registered topics. This feature in turn allows us to distinctively regionalize clusters based on their population, and allocate brokers to each region, while the brokers are connected, and be able to perform the role management with ease. In such an establishment, the clients need only to connect to their region's broker and contribute to the entire FL system, ensuring that their contribution can reach other regions as well.


\section{Implementation}
The core of SDFLMQ is implemented using Python. MQTTFC, the base of SDFLMQ for remote function calls, is also implemented using Python and the EMQX Paho client class which provides an interface to establish Pub/Sub communication. To manage large message payloads, such as a set of parameters in a deep neural network, a batching mechanism is implemented at the core of MQTTFC, which serializes the payload and divides it into multiple batches before sending. The batches are encoded also and batch\_ids are allocated to them before transmission. As the batches are received, they are decoded, and compiled at the receiver site. Messages are also sent in customized separable text format, while session stats and cluster topologies are encoded into JSON format. For larger payloads, a compression mechanism using \textit{zlib} library is implemented at the core of the SDFLMQ\_Client code. \textit{PSUtil} and \textit{Tracemalloc} libraries are also used to collect system stats such as available memory, bandwidth, and CPU clock cycles. SDFLMQ is openly accessible to the public here \cite{SDFLMQsource}

\section{Use-Cases}

Earlier we mentioned the importance of the lightweight implementation feature of our SDFLMQ framework. Here we show in a code snippet that in fact SDFLMQ can be invoked and integrated into a training pipeline with only a few lines of code. The code in Listing \ref{code1} is a pipeline in which a fully connected multi-layer perceptron (MLP) is initialized and trained with local data for 5 epochs, and sent to an aggregator for global model updating. As shown, The only parts where SDFLMQ-related codes are operated are the parts where a federated learning client is established (line 12), a session is created (line 19), a local model is sent for global updating after local training is finished (line 50, 51, 52).


\begin{lstlisting}[language=Python, label=code1, caption=Code snippet of a client creating a session to perform 2 federated learning rounds and optimize an MLP model for handwritten digit detection using MNIST dataset.]
from Core.sdflmq_client_logic import SDFLMQ_Client

# Setup local training parameters
train_dataset = get_MNIST()
model = get_MLP()
criterion = nn.CrossEntropyLoss()
optimizer = optim.Adam(model.parameters(), lr=0.001)

# Setup SDFLMQ client
FL_ROUNDS = 2
myid = "client_" + str(random.randint())         
fl_client = SDFLMQ_Client(  myID=myid,
    broker_ip = 'localhost',
    broker_port = 1883,
    preferred_role="aggregator",
    loop_forever=False)
    
# USE CODE BELOW TO CREATE A SESSION:
fl_client.create_fl_session(session_id="session_01",
    session_time=timedelta(hours=1),
    session_capacity_min= 5,
    session_capacity_max= 5,
    waiting_time=timedelta(minutes=2),
    fl_rounds=FL_ROUNDS,
    model_name="mlp",
    preferred_role = 'aggregator')

# USE CODE BELOW TO JOIN A SESSION:
# fl_client.join_fl_session(session_id="session_01",
#     fl_rounds=FL_ROUNDS,
#     model_name="mlp",
#     preferred_role = 'aggregator')

# Optimization Loop
for round in range(FL_ROUNDS):    
    num_epochs = 5
    for epoch in range(num_epochs):
        # Local training
        model.train()
        for images, labels in train_dataset:
            images, labels = images, labels
            # Forward pass
            outputs = model(images)
            loss = criterion(outputs, labels)
            # Backward pass and optimization
            optimizer.zero_grad()
            loss.backward()
            optimizer.step() 
    # Federated learning
    fl_client.set_model('session_01',model)
    fl_client.send_local('session_01')
    fl_client.wait_global_update()

\end{lstlisting}

\section{Runtime Evaluation}

To evaluate the performance of SDFLMQ, we executed multiple scenarios with various compositions. At this early stage of evaluation, we only targeted changing the number of contributing clients, as well as the topology of their clustering. We ran a pipeline to train and optimize a fully connected MLP model to detect handwritten digits using the well-known MNIST dataset. While the MNIST dataset typically includes 60,000 training samples, we aimed to include only a fraction of the training dataset of which we specify per scenario.

\begin{figure}[t]
    \centering
    \includegraphics[width=1\linewidth]{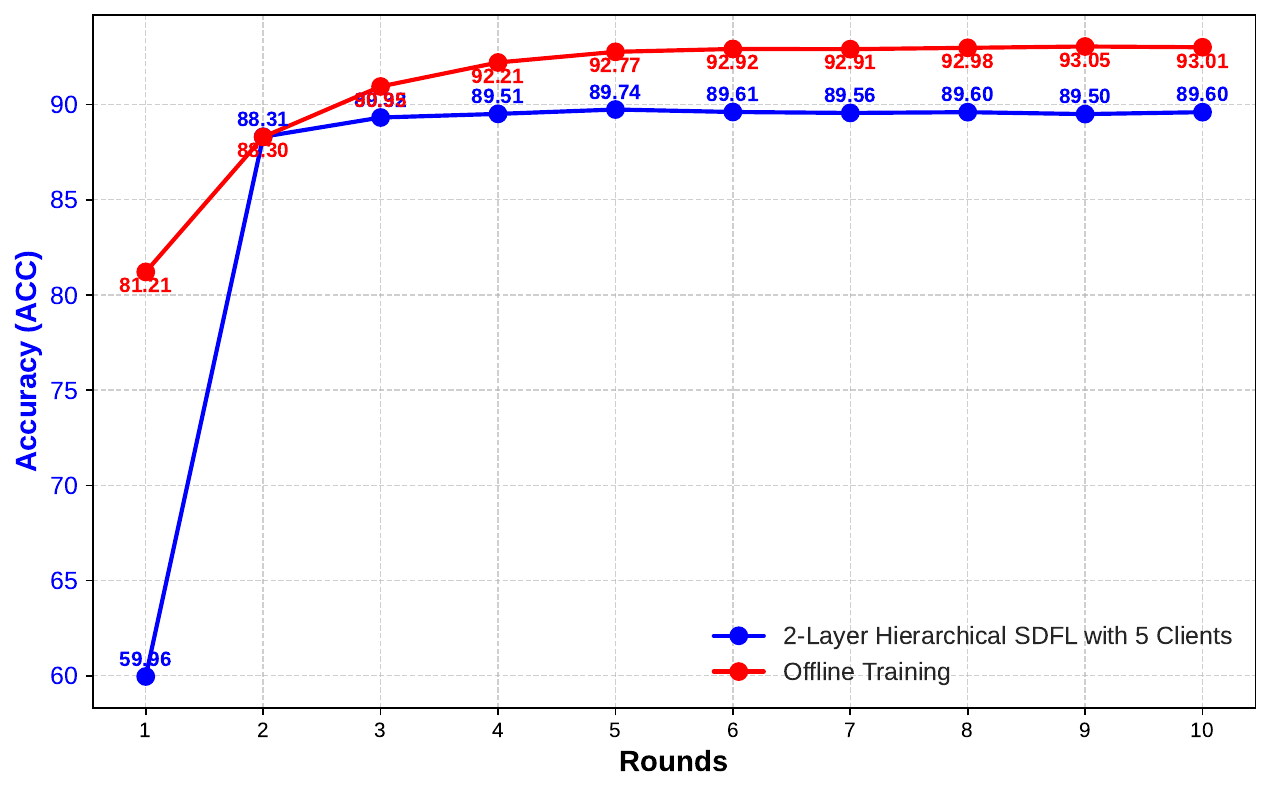}
    \caption{Comparing MLP model accuracy convergence in Offline training with training using SDFLMQ via 5 clients.}
    \label{fig:acc_1}
\end{figure}

The first evaluation we ran was to compare the convergence rate of the prediction accuracy (testing) of local training with federated learning wherein 5 clients contribute to the global model update process. In the federated learning part of the scenario, we gave each client $1\%$ of the MNIST dataset. For the aggregation, we also defined using FedAvg, which is one of the typical aggregation methods in FL. To set an equal ground, in the local training part of the scenario, we gave only $5\%$ of the dataset to the training pipeline. Fig. \ref{fig:acc_1} shows the convergence rate of the two cases, and we can see that the FL case converges to a value close to what the local-training case is converging which is around $90\%$. 

\begin{figure}[t]
    \centering
    \includegraphics[width=1\linewidth]{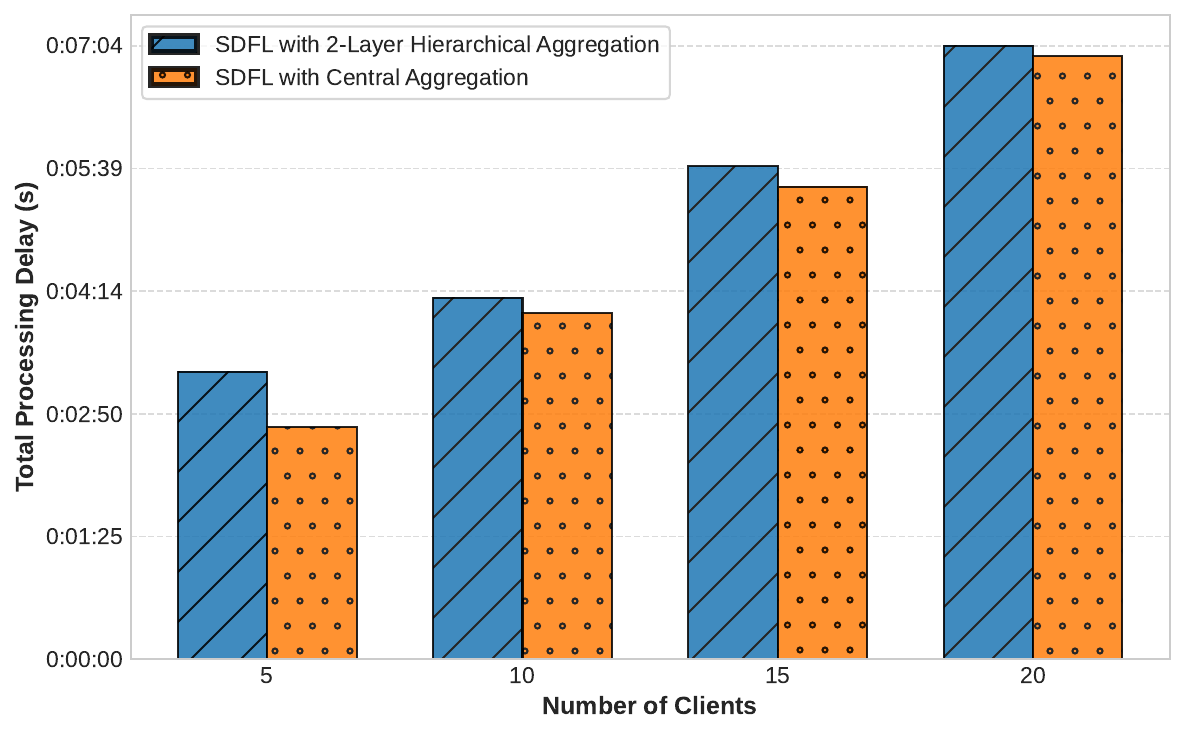}
    \caption{Total processing delay of running 10 FL rounds with two different aggregation topologies.}
    \label{fig:performance}
\end{figure}

In the next evaluation, we set to analyze the performance of SDFLMQ in terms of the total processing delay while we gave it different policies for clustering. In this evaluation, we tested two scenarios: In one case, we set the framework to cluster the nodes in a tree-like topology, forming hierarchical clusters. In specific, we defined 3 layers in the hierarchy, wherein the first layer includes a root aggregator unit, the second layer includes the intermediate aggregator units, and the third layer includes the training-only units. We set the number of aggregator units proportional to the total number of clients, wherein $30\%$ of the clients would be aggregators. In the second case, we set the framework to create only one cluster and set only one unit as an aggregator. We then tested the two cases for various numbers of contributing clients. Fig. \ref{fig:performance} shows the total processing delay of global model updating with respect to different numbers of contributing clients. An interesting take here is the difference of the two cases is not as significant, even though in the 2-layer Hierarchical SDFL we have more than one level of aggregation. Nonetheless, one can also notice that with the growing number of contributing clients, the gap between the two cases is also closing further. While it is too early to draw a definitive conclusion, this can at least be potential evidence that relying only on one unit for aggregation can induce further delay if number of contributing clients is large. That could be due to various systematic factors, such as bandwidth overload or memory overload, which needs further in-depth studies.


\section{Future Expansions of SDFLMQ}
SDFLMQ is set to be a publicly available framework to practice federated learning at the edge. However, while the framework is already set as an open-source project, several key expansions need to be developed to increase the efficiency of the framework and its applicability for constrained systems at the edge. These potential expansions are:
\begin{itemize}
    \item Dynamic Aggregation placement via swarm intelligence optimization and genetic algorithm. Integrating such methods for optimization, especially as a black-box optimizer, can play key parts in optimizing SDFLMQ operations with zero reliance on application-specific information, and solely on the performance of the framework in delivering the global models to the client machines.
    \item Dynamic Clustering using non-parametric supervised learning methods. So far the clustering in SDFLMQ follows fixed policies and procedures. Nonetheless, we anticipate that adjusting the cluster topologies per round, similar to the aggregation placement, can further optimize the performance of the framework. 
\end{itemize}

\section{Conclusion}
In this paper, we proposed a new framework for federated learning that allows the governance of the process without relying on a central server. We proposed SDFLMQ, a framework for semi-decentralized federated learning over MQTT communication protocol, which has two key features including clustering and dynamic aggregation placement. We discussed how MQTT, and in general any publish/subscribe communication protocol can help deliver a facilitated and dynamic medium for clustering and role association using the topic-based message passing. We showed that we used this mechanism SDFLMQ to associate the aggregation role to clients, which can vary per federated learning round with negligible communication cost. We performed some primary evaluations of our framework and showed that it can deliver on par with what a central federated learning can in terms of processing delay, and on par with what a local training pipeline can in terms of model accuracy.

\bibliographystyle{ieeetr}
\bibliography{main}

\end{document}